\documentclass[a4paper,11pt]{article}
\pdfoutput=1
\usepackage{jheppub} 
\usepackage[T1]{fontenc} 
\usepackage[table]{xcolor}
\usepackage[caption=false]{subfig}
\usepackage{epsfig}
\usepackage{colortbl}
\usepackage[T1]{fontenc} 
\usepackage{tikz-feynman}
\usepackage{hepnames}
\usepackage{amsmath}
\usepackage{amssymb}
\usepackage{adjustbox}
\usepackage{multirow}
\usepackage{slashed}
\usepackage{cleveref}
\usepackage{soul}
\usepackage{xcolor}
\usepackage[font=small]{caption}
\usepackage[font={small},labelfont={bf}]{caption}
\usepackage{url}
\definecolor{MyDarkBlue}{rgb}{0.1, 0.1, 0.8}
\definecolor{SBlue}{rgb}{0.2, 0.4, 0.7} 
\definecolor{MyLightBlue}{rgb}{0.22,0.51,0.9}
\definecolor{MyGreen}{rgb}{0.0, 0.5, 0.0}
\definecolor{BrickRed}{rgb}{0.8, 0.25, 0.33}
\usepackage{hyperref}
\hypersetup{colorlinks, citecolor=BrickRed,linkcolor=MyDarkBlue, urlcolor=MyGreen}
\title{Direct Search signal of two-component Dark Matter}
\author[a]{Subhaditya Bhattacharya, }
\author[a]{Dipankar Pradhan.}
\affiliation[a]{Department of Physics, Indian Institute of Technology Guwahati,\\
North Guwahati, Assam-781039, India.}
\emailAdd{subhab@iitg.ac.in}
\emailAdd{d.pradhan@iitg.ac.in}
\abstract{How do we know if the dark sector consists of more than one dark matter (DM) component is an important question, for which the answer is not very definite. 
In this article we study such a possibility in context of direct DM search. It was pointed out earlier in a model independent analysis that a kink in the nuclear 
recoil energy spectrum may indicate to the presence of two DM components. However, realising one such model was difficult due to experimental constraints. 
Here we propose and study a model containing a vector boson DM and a scalar DM, aided by a light scalar mediator, where a kink in the 
nuclear recoil spectrum arises after addressing individual relic densities, direct search limits, collider constraints and theoretical limits. 
We find out the allowed parameter space of the model and those regions likely to show such distinctive signal.}
\begin{document}
\makeatletter
\gdef\@fpheader{}
\makeatother
\maketitle
\flushbottom
\section{Introduction}
\label{introduction}
Albeit the hints from astrophysical and cosmological observations that our observed universe contains 26.8\% \cite{Planck:2018vyg, Bertone:2004pz} dark matter (DM), its 
detection is yet to be achieved. Apart from the stability, massive nature and electromagnetic charge neutrality, we remain largely nescience about DM characteristics. 
One particular question is often asked whether the dark sector consists of a single particle or many, like the visible sector does. And if so, then how the observations 
related to relic density, and search strategies alter. This is one of the points of study in this paper.  

When the DM freezes out of thermal bath, gives rise to WIMP \cite{Cline:2013gha,Roszkowski:2017nbc,Lee:1977ua,Gondolo:1990dk} or SIMP \cite{Hochberg:2014dra} type candidates, while FIMP \cite{Hall:2009bx} is produced non-thermally. Apart, several possibilities exist. In multicomponent frameworks, DM-DM interaction is 
constrained from galaxy cluster observations \cite{Clowe:2003tk, Markevitch:2003at, Randall:2008ppe, Kahlhoefer:2015vua,10.1111/j.1365-2966.2011.19266.x}. WIMP-WIMP \cite{Profumo:2009tb, Bhattacharya:2013hva} scenario has best detectability as WIMP has sizeable DM-SM interaction, while WIMP-FIMP \cite{DuttaBanik:2016jzv}, FIMP-FIMP \cite{Pandey:2017quk, PeymanZakeri:2018zaa}, SIMP-SIMP \cite{Ho:2022erb, Choi:2021yps}, WIMP-pFIMP\footnote{Pseudo-FIMP (pFIMP) arises in a multi-component DM frameworks, having feeble connection to SM, but adequate interaction with the partner DM.} \cite{Bhattacharya:2022vxm, Bhattacharya:2022dco, Bhattacharya:2024nla, Lahiri:2024rxc}, SIMP-pFIMP \cite{Bhattacharya:2024jtw} frameworks offer plenty of phenomenology. 

DM search mainly relies on the DM-SM interaction and is done mainly via three strategies, direct, indirect and collider searches. With minimal electromagnetic coupling, 
DM misses the detector, and produces missing momenta or energy. We cite some example signal/background analysis at upcoming colliders, like HL-LHC \cite{ZurbanoFernandez:2020cco}, FCC \cite{Tomas:2018wvi}, CEPC \cite{CEPCStudyGroup:2023quu}, muon collider \cite{Accettura:2023ked}, ILC \cite{Bambade:2019fyw} etc. Occasionally, the collider may also probe the mediator for the interaction between DM and SM, see \cite{Buchmueller:2017qhf}. Non observation of DM at collider put bounds on 
DM-SM interaction and mass, the ones relevant include Tevatron \cite{CDF:2012wzv}, LEP \cite{LEP:1991hsu} and LHC \cite{Evans:2008zzb}.
Indirect detection (ID) relies on excess of SM particles (e.g., charged leptons, gamma rays, neutrinos, protons, anti-particles etc.) produced by the decay or annihilation of DM particles \cite{Gaskins:2016cha} in regions with high DM density, such as the Galactic Center and dwarf spheroidal galaxies. Indirect searches provide constraints on DM self-annihilation and semi-annihilation, for eg., Fermi-LAT \cite{Fermi-LAT:2015att, Fermi-LAT:2016afa}, H.E.S.S \cite{HESS:2016mib}, and CTA \cite{Silverwood:2014yza}.
In direct detection (DD), DM scatters off the detector nuclei or atomic electrons (depending on mass), 
and results in recoil of the nuclei/electrons producing scintillation signal in the detector \cite{Goodman:1984dc, Essig:2011nj}.  
As our focus here is on cold DM in the GeV-TeV mass range, which plays a crucial role in supporting the universe's structure formation, DM produces nuclear recoil. The
most relevant constraints in such cases are derived from experiments such as PandaX-xT \cite{PANDA-X:2024dlo}, XENONnT \cite{XENON:2023cxc}, LUX-ZEPLIN \cite{LZCollaboration:2024lux}, and XLZD \cite{XLZD:2024nsu} amongst others. Heavy DM search is also underway with XENONnT \cite{XENON:2023iku} and sub-GeV DM searches with SuperCDMS \cite{SuperCDMS:2020aus, SuperCDMS:2020ymb, SuperCDMS:2023sql}, EDELWEISS \cite{EDELWEISS:2020fxc}, CRESST-III \cite{CRESST:2022lqw}, DarkSide-50 \cite{DarkSide:2022dhx}, CDEX-10 \cite{CDEX:2022kcd}, DAMIC-M \cite{DAMIC-M:2023hgj}, SENSEI \cite{SENSEI:2023zdf}, etc.

In a model-independent analysis of two-component DM setup \cite{Herrero-Garcia:2017vrl, Herrero-Garcia:2018qnz},\footnote{In context of colliders, references \cite{Bhattacharya:2022qck, Bhattacharya:2022wtr} have shown that a two-component DM scenario can lead to the appearance of two peaks in the missing energy ($\slashed{E}$) distribution; reference \cite{Cao:2007fy} attempted to explore how the indirect signal (specifically the photon flux) is altered in the presence of two DMs.} it was shown that a kink may appear in the recoil energy spectrum of DD, when one DM is of low mass $\sim 10~\rm GeV$, and the other is heavy, $\gtrsim 40~\rm GeV$.
The only physical observable in the event rate analysis is the curvature of the event rate spectrum, which would serve as evidence for the presence of more than one DM particle in the dark sector. The position and properties of this curvature depend not only on the DM masses but also on local DM densities and the DM scattering rates with detector nuclei.
The study motivated us to look into a UV complete set up where such observations can be realised. This is challenging as having a low mass WIMP faces severe constraint from 
experimental data, particularly invisible decay constraints of Higgs or the DM-SM portal. Further, the relative relic densities and DM masses are 
closely related in a realistic model, so that finding a parameter space where such novel feature can be seen in future sensitivities of DD becomes an interesting exercise.

In this work, SM is extended by a gauge singlet vector boson DM (VBDM), which acquires mass through the spontaneous breaking of a $U(1)_X$ symmetry via a complex scalar. Apart we postulate the presence of a real scalar DM (RSDM). The stability of both is ensured by an appropriate $\mathcal{Z}_2 \otimes \mathcal{Z}_2^{\prime}$ symmetry. Through the mixing with CP-even part of the SM Higgs doublet, a new real scalar particle emerges along with the SM Higgs \cite{ATLAS:2023oaq, Biekotter:2023jld}, playing a crucial role 
in the analysis. Our goal is to study the allowed parameter space satisfying relic density, DD, ID, and the collider constraints, that result in a visible kink in the total event rate spectrum indicating the presence of more than one DM components.

This paper is organized as follows: in Section\,.~\ref{mod:-vectro-scalar}, we discussed the possibility of getting two stable vector and scalar DM components transforming under a single $\mathcal{Z}_2\otimes\mathcal{Z}_2^{\prime}$ symmetry and in Section\,.~\ref{sec:cbeq-relic} we have discussed the coupled Boltzmann equation and its solution- relic density of DMs. The direct detection signal for this vector-scalar DM model is discussed in Section\,.~\ref{sec:2dm-dd}. We finally summarise in Section\,.~\ref{sec:summary}. Appendix \ref{sec:feyn}, and \ref{sec:equib} provide some necessary details omitted in the main text.
\section{The Model}
\label{mod:-vectro-scalar}
This article aims to come up with a UV-complete model that produces a visible kink in the time-averaged event rate spectrum while satisfying all the 
current theoretical and observational constraints. Its worth mentioning that all the simple minded two component DM models, 
consisting of singlet or doublet scalar or fermion, leaves in a mass regime where such distinguishability is difficult to address. 

To alleviate this, we extend the SM gauge symmetry by a global $U(1)_X$ symmetry under which the 
complex scalar $S$ is charged. The associated dark gauge boson $X_{\mu}$ is stabilized by a $\mathcal{Z}_2$ symmetry, making it a viable vector boson DM candidate.  
We further include a real scalar $\phi$ singlet, stable under a $\mathcal{Z}_2^{\prime}$ symmetry, representing a second DM candidate. The charges of these particles under the $\mathcal{Z}_2 \otimes \mathcal{Z}_2^{\prime}$ symmetry are listed in tab\,.~\ref{tab:tab1}.
\begin{table}[htb!]
\begin{center}
\begin{tabular}{|c|c c|}\hline
\rowcolor{gray!30}\bf Dark Fields&  \multicolumn{2}{c|}{$\mathcal{Z}_2\otimes \mathcal{Z}_2^{\prime}$}\\
\rowcolor{cyan!10}Real scalar $\phi$&$~\phi$&$-\phi$\\
\rowcolor{green!10}Complex scalar $S$&$~S^*$&$~S$\\
\rowcolor{lime!10}$U(1)_X$ Gauge Boson $X_{\mu}$&$-X_{\mu}$&$~X_{\mu}$\\\hline
\end{tabular}
\end{center}
\caption{Model particles and their charges under $\mathcal{Z}_2\otimes\mathcal{Z}_2^{\prime}$ symmetry.}
\label{tab:tab1}
\end{table} 
The SM extended dark sector Lagrangian is written as \cite{Bhattacharya:2021rwh},
\begin{align}
\mathcal{L}=\mathcal{L}_{\rm SM}+|\partial_{\mu}\phi |^2+|D_{\mu}S|^2+\dfrac{1}{4}X_{\mu\nu}X^{\mu\nu}-{ V(\phi,S,H)}\,,
\end{align}
 where,
 $$
 D_{\mu}=\partial_{\mu}+ig_X X_{\mu};~~~X_{\mu\nu}=\partial_{\mu}X_{\nu}-\partial_{\nu}X_{\mu};
 $$
 and,
\begin{align}\nonumber
V(\phi,S,H)=&-\mu_H^2(H^{\dagger}H)+\lambda_H(H^{\dagger}H)^2+\dfrac{1}{2}\mu^2_{\phi}\phi^2+\dfrac{1}{4!}\lambda_{\phi}\phi^4-\mu_S^2(S^*S)+\lambda_S(S^*S)^2\\&
+\dfrac{1}{2}\lambda_{\phi H}\phi^2(H^{\dagger}H)+\dfrac{1}{2}\lambda_{\phi S}\phi^2(S^*S)+\lambda_{HS}(H^{\dagger}H)(S^*S)\,.
\label{eq:V}
\end{align}
The potential $ V(\phi,S,H)$ in eq\,.~\ref{eq:V}, for $\lambda_{H},\lambda_S,\lambda_{\phi},\mu_{\phi}^2>0$ and $\mu_H^2<0,\mu_S^2<0$, so that it provides the following vacuum:
\begin{align}
H=\begin{pmatrix}
\phi^+ & \\
\frac{v+h+i\phi_0}{\sqrt{2}} &\\
\end{pmatrix}  \to 
\langle H \rangle =\begin{pmatrix}
0 & \\
\frac{v}{\sqrt{2}} &\\
\end{pmatrix}\,; 
\end{align}
\begin{align}
S =\frac{v_s+s+i\mathcal{A}}{\sqrt{2}} \to \langle S \rangle =\frac{v_s}{\sqrt{2}}; 
~\langle \phi \rangle=0\,. 
\end{align} 
 In the above, $\phi^{\pm,0},~\mathcal{A}$ denote Nambu-Goldstone Bosons \cite{Goldstone:1962es,Nielsen:1975hm,Coleman:1973ci,Burgess:1998ku}, which disappear in the unitary gauge after EWSB. The breaking of $U(1)_X$ makes the associated gauge boson massive:
\begin{align}
m_X=v_sg_X\,,
\end{align}
where $g_X$ denotes $U(1)_X$ gauge coupling constant and $v_s$ denotes the $U(1)_X$ symmetry breaking scale. Let us have a quick look into the relevant constraints.
\section*{$\bullet$ Tree level unitarity}
The tree-level unitarity of the theory comes from all possible $2\to 2$ scattering amplitude and can be ensured by \cite{Bhattacharyya:2015nca, Horejsi:2005da},
\begin{align}
\lambda_{H}<4\pi,~\lambda_{S}<4\pi,~\lambda_{HS}<8\pi\,.
\end{align}
\section*{$\bullet$ Perturbativity}
To maintain the perturbativity of the theory, the couplings of the theory obey \cite{Bhattacharya:2019fgs},
\begin{align}
\lambda_H<4\pi,\quad \lambda_S< 4\pi,\quad g_X<\sqrt{4\pi},\quad \lambda_{HS}<4\pi \,.
\end{align}
\section*{$\bullet$ Limits on the thermal Dark Matter mass}
There is a unitarity bound that sets an upper limit on the mass of symmetric and asymmetric thermal DM, is $\sim ~110~\rm TeV$ \cite{Griest:1989wd, Baldes:2017gzw, Smirnov:2019ngs, Bhatia:2020itt}. The combined Big Bang Nucleosynthesis (BBN) and Cosmic Microwave Background (CMB) provides the lower limits to the WIMP masses is $\sim 0.5-5~\rm MeV$ \cite{Zhou:2022ygn, Krnjaic:2019dzc, Nollett:2013pwa, Depta:2019lbe, Serpico:2004nm, Ho:2012ug, Boehm:2013jpa, Nollett:2014lwa, Sabti:2019mhn}.
\section*{$\bullet$ Relic density}
The Planck data \cite{Planck:2018vyg} constrains the present DM relic density,
\begin{align}
\rm\Omega_{DM}h^2=0.1200\pm 0.0012\,.
\end{align}
where $h$ is the reduced Hubble parameter $\rm H_0/(100~\rm km/s/ Mpc)$ with $\rm H_0 = 67.4\pm 0.5~ km/s/Mpc$ being the current Hubble constant.
\section*{$\bullet$ Ranges of mixing angle $\sin{\vartheta}$}
As the scalar mass matrix is real, symmetric and non-diagonal, it can be diagonalised by an orthogonal matrix, resulting in mass eigenstates $(h_1,h_2)$.
\begin{gather}
\begin{pmatrix}
h_1  \\ h_2 
\end{pmatrix}=\begin{pmatrix}
\cos\vartheta &-\sin\vartheta \\
\sin\vartheta &\cos\vartheta
\end{pmatrix}\begin{pmatrix}
h  \\ s
\end{pmatrix}\,.
\end{gather}
Here we assume $h_1$ to be SM like, and $h_2$ is lighter.  The relations between the parameters in the Lagrangian, to the physical ones and mixing are,
\begin{gather}
 \mu_H^2 =  (\lambda_H v^2+\frac{1}{2}\lambda_{HS}v^2_s)\,, \\
 \mu_S^2 = (\lambda_S v^2_s+\frac{1}{2}\lambda_{HS}v^2)\,, \\
 \lambda_{H S} = \frac{\sin2\vartheta}{2v_sv}\left(m^2_{h_2}-m^2_{h_1}\right)\,,\\
 \lambda_H = \frac{1}{2v^2}\left(m^2_{h_1}\cos^2\vartheta+m^2_{h_2}\sin^2\vartheta\right)\,,\\
  \lambda_S = \frac{1}{2v_s^2}\left(m^2_{h_2}\cos^2\vartheta+m^2_{h_1}\sin^2\vartheta\right)\,.
\label{}
\end{gather}
The Higgs precision measurement set an upper limit on the $h-s$ mixing angle $\vartheta$ at $95\%~\rm CL$ for $\rm125.1~GeV$ Higgs \cite{ATLAS:2015ciy, Tuominen:2021wrl, Chalons:2016lyk, Robens:2015gla, Chalons:2016jeu, Robens:2022zav, Robens:2023bzp,Lane:2024vur},
\begin{align}
|\sin\vartheta|\lesssim0.29\,.
\end{align}
In our analysis, we have chosen a small but not overly tiny mixing angle ($10^{-4}\lesssim|\sin\vartheta|\lesssim10^{-2}$), ensuring $h_2$ stays in thermal equilibrium during DM freeze-out, has a lifetime shorter than the BBN era ($\tau_{h_2}<1 ~\rm sec$), and does not disrupt BBN or induce invisible Higgs decays.
\section*{$\bullet$ Higgs invisible decay}
If DM's connected via Higgs portal, have smaller mass than half of the SM Higgs, then Higgs can invisibly decay to DM pair. The most sensitive limits on 
$\mathcal{B}_{h_1\to \rm inv}$ are obtained in $\rm VBF$ searches exploring data collected at $\sqrt{s} = 13\rm ~ TeV$, 
excluding $\mathcal{B}_{h_1\to \rm inv} <0.18~(0.10)~\rm$ observed (expected) at $95\%$ C.L using 138 $\rm fb^{-1}$ of CMS data,
\cite{CMS:2022qva}, and $\mathcal{B}_{h_1\to \rm inv} < 0.15 ~(0.10)$ using 139 $\rm fb^{-1}$ of ATLAS data \cite{ATLAS:2022yvh}.
Together, searches for invisible decays of the Higgs boson using $139~ {\rm fb}^{-1}$ of $p~p$ collision data at
$\sqrt{s}=13~\rm TeV$ recorded in Run 2 data of LHC set an upper limit on the invisible Higgs boson branching
ratio of $\mathcal{B}_{h_1\to \rm inv}<0.113~(0.080_{-0.022}^{+0.031})$ observed (expected) at the $95\%$ CL. 
Here we assume that the dark Higgs $h_2$ decays instantaneously inside the detector, with DMs long-lived. 
In our context, therefore, the invisible decay width of the SM Higgs is,
\begin{align}
\rm\Gamma_{h_1}^{ inv}=\dfrac{\mathcal{B}_{h_1\to  inv}}{1-\mathcal{B}_{h_1\to  inv}}\Gamma_{h_1}^{ SM}\,,
\end{align}
where, $\rm\Gamma_{h_1}^{ inv}=\Gamma_{h_1\to XX}+\Gamma_{h_1\to h_2h_2}$ and $\Gamma_{h_1}^{\rm SM}=4.100 \times 10^{-3}~\rm GeV~(\pm1.4\%)$ \cite{LHCHiggsCrossSectionWorkingGroup:2016ypw} for $m_{h_1}=125.09~\rm GeV$. 
\section*{$\bullet$ Can $h_2$ be a 95 GeV scalar?}
The latest CMS analysis confirms an excess of di-photon events around 95 GeV. By combining data from the first three years of Run 2, collected at $\sqrt{s}=13\rm~TeV$ with integrated luminosities of $36.3~\rm fb^{-1}$, $41.5~\rm fb^{-1}$, and $54.4~\rm fb^{-1}$, CMS reports a local (global) significance of $2.9~(1.3)\sigma$ at a mass of 
$\rm 95.4~GeV$ \cite{CMS:2024yhz}.
In contrast, ATLAS reported di-photon search results below $125\rm~GeV$ using $80~\rm fb^{-1}$ of Run 2 data \cite{ATLAS:2023jzc} to show only a mild excess, 
with the largest deviation at $\rm 95.4~GeV$, yielding a local significance of $1.7\sigma$ \cite{ATLAS:2024bjr}.
A combined analysis gives a signal strength of $\mu_{\gamma\gamma}^{\rm ATLAS+CMS}=0.24^{+0.09}_{-0.08}$ \cite{Biekotter:2023oen}, corresponding to a 
$3.1\sigma$ excess. Some studies explored the possibility with $ U(1)_X$ gauge boson DM model, 
containing a real scalar $ h_2 $ along with the SM Higgs $h_1$ \cite{YaserAyazi:2024hpj}, 
investigating the processes $ q\overline{q} \to t\overline{t}X_{\mu}X^{\mu} $ and $ q\overline{q} \to VX_{\mu}X^{\mu} \) with $ V \to \ell\overline{\ell} $, with 
$ m_{h_2} = 95.4~{\rm GeV} $ and $ V = Z_{\mu}, h_1 $. No exclusions apply to our region of interest, provided that $g_X<10^{-2}$ and $|\sin\vartheta|<10^{-2}$.
The reference \cite{Gao:2024ljl} provides theoretical predictions for signal strengths in $\gamma\gamma$, $\tau\overline{\tau}$, and $b\overline{b}$ channels in the $ U(1)_X \text{SSM} $ model, which aligns well with the excesses observed by CMS. Interestingly, LEP data strongly disfavours the production of 95 GeV scalar particle, 
as well as any other new physics interpretation in the 95-100 GeV mass range \cite{Janot:2024ryq}. We do not intend to perform a full analysis of $h_2$ to diphoton decay, which can cause this excess near $95~\rm GeV$.
\section*{$\bullet$ Bounds on light mediator $h_2$ mass from thermalization condition}
The decay or annihilations of the light scalar $h_2$ into SM fermions is restricted by BBN data, because the precision measurement of the baryon density through 
BBN and CMB are well measured \cite{Steigman:2012ve, Kaplinghat:2013yxa, Belanger:2022qxt} and restricts $h_2$ lifetime $\rm\tau_{h_2}\lesssim 1~sec$ \cite{Fradette:2017sdd, Pospelov:2010hj}. The $h_2$ decay life time in our case turns,
\begin{align}
\rm\tau_{h_2}=\left[\Gamma_{h_2\to SM~SM}+\Gamma_{h_2\to X~X}+\Gamma_{h_2\to\phi~\phi}\right]^{-1}\,.
\end{align}
Equivalently, the total decay width of $\rm h_2~(\Gamma_{h_2})$ should be greater than $6.58 \times 10^{-25} \, \rm GeV$, which puts a stringent limit on the model parameters.
For other relevant constraints for the light scalar $h_2$ in presence of VBDM see \cite{Duch:2017khv}. Also, you may visit ref\,.~\cite{Clarke:2013aya, Schmidt-Hoberg:2013hba} for relevant bounds on light scalar DM from LEP, LHC, light meson decay, and fixed target experiments.
For $h_2$ to remain in chemical equilibrium with thermal bath, the interaction rate of $\rm h_2$ must be larger than the Hubble expansion rate,
\begin{align}
\sum_{\mathcal{F}} \left[{\rm\langle\Gamma\rangle_{h_2\to \mathcal{F}~\mathcal{F}}+n_{h_2}^{ eq}}\langle \sigma  v\rangle_{\rm h_2~h_2\to\mathcal{F}~\mathcal{F}}\right]_{\rm T^{DM}_{FO}}\gtrsim \rm H(\rm T^{ DM}_{FO})\,,
\end{align}
where $\rm \mathcal{F}\in\{SM~particles,~\phi,~X\},~and~ T^{DM}_{FO}\sim m_{DM}/25$. In fig\,.~\ref{fig:thermal-h2}, see the the variation of interaction rate of $h_2$ with the bath temperature. However, the kinematic equilibrium of $h_2$ is maintained by the inelastic scattering with light SM fermions $(\rm h_2~\mathcal{F}\to h_2~\mathcal{F})$, sharing the same SM bath temperature ($\rm T$), so that we use $\rm n_{h_2}(T)\simeq n_{h_2,0}^{}(T)$ throughout the analysis; one can also study the departure from the aforementioned condition via solving the three coupled Boltzmann equations (cBEQ) simultaneously.
\section*{$\bullet$ Limits on thermal dark matter annihilation}
CMB anisotropies constrain energy injection from DM annihilation, providing limits on WIMP annihilation cross-sections that complement indirect DM searches. 
Planck 2018 data constrains DM mass and annihilation cross-section; the strongest bounds are obtained assuming s-wave annihilation into bottom quark pairs, $\langle\sigma v\rangle_{\rm DM~DM\to b\overline{b}}\sim 8\times 10^{-27}\rm cm^3/s$ for $\rm m_{DM}^{}= 6~GeV$ \cite{Planck:2018vyg}. On the other hand, the 95\% confidence level upper limits on the thermally-averaged cross-section for DM particles annihilating into $b\overline{b}$, derived from a combined analysis of 158 hours of Segue 1 observation with MAGIC, and 6-year observations of 15 dwarf satellite galaxies by the Fermi-LAT, is $ \sim 10^{-26} \, \rm{cm}^3/\text{s} $ for $\rm m_{DM}^{}= 6~GeV$ \cite{MAGIC:2016xys}.
\section{Coupled Boltzmann Equation and Relic density}\label{sec:cbeq-relic}
In this study, we have considered both the dark sector particles as WIMP-like DM, so that both components can be probed in future DD experiments. In particular, we are interested in
the possibility of getting a kink in the recoil energy spectrum. But before that, we will address the possibility of the DM components to acquire correct relic density. 
WIMPs are initially in thermal equilibrium with the bath particles and undergo freeze-out when their interaction rate falls below the Hubble expansion rate.
The Feynman diagrams, which provide the annihilation of DMs and DM-DM conversion processes, are shown in fig\,.~\ref{fig:feynman}.
The freeze out of DM components are evaluated by solving the coupled Boltzmann Equations (cBEQs), which in our case, reads:
\begin{eqnarray}
\begin{split}\centering
\dfrac{dY_{\phi}^{}}{dx}=-\frac{\bf{s}}{x~H(x)}&\left[\left(Y_{\phi}^2-Y_{\phi,0}^{2}\right)\langle\sigma v\rangle_{\phi~\phi\to \rm SM~SM}\,+\,\left(Y_{\phi}^2-Y_{X}^2\frac{Y_{\phi,0}^{2}}{Y_{X,0}^{2}}\right)\langle\sigma v\rangle_{\phi~\phi\to X ~X}\right]\,,\\
\dfrac{dY_{X}^{}}{dx}=-\frac{{\bf s}}{x~H(x)}&\left[\left(Y_{X}^2-Y_{X,0}^{2}\right)\langle\sigma v\rangle_{ \rm X~X\to SM ~SM} \,+\,\left(Y_X^2 - Y_{\phi}^2 \dfrac{Y_{X,0}^{2}}{Y_{\phi,0}^{2}}\right)\langle \sigma v\rangle_{XX\to\phi~\phi}\right]\\\,
+\,\frac{2}{x~H(x)}\sum_{i=1}^2&\left[Y_{h_i,0}^{}-Y_{h_i,0}^{}\dfrac{Y_X^2}{Y_{X,0}^{2}}\right]\langle\Gamma\rangle_{h_i\to X~X}\,.
\label{eq:cbeq_model}
\end{split}
\end{eqnarray}
In the above, notations have very standard meaning, some of them worth noting are, $x=\mu_{\phi X}/T$, where $\mu_{\phi X}^{}=m_{\phi}m_X/(m_{\phi}+m_X)$ 
denotes reduced mass for the two DM components, $T$ denotes temperature of the thermal bath, 
$Y_{i,0}^{}=n_{i,0}^{}/{\bf s}$ denotes equilibrium yield, with ${\bf s}=\frac{2\pi^2}{45}g_*^s T^3$ denoting entropy density, 
$n^{}_{i,0}=\frac{T}{2\pi^2}g_im_i^2K_2\left(\dfrac{m_i}{T}\right)$ indicate to equilibrium number density, annihilation final states include ${\rm SM \in \{h_1,~h_2,~W^{\pm},~Z,~leptons ~and~quarks\}}$, and
\begin{align}
\langle\sigma v\rangle_{\rm a~b\to c~d}~n_{a,0}n_{b,0}=\langle\sigma v\rangle_{\rm c~d\to a~b}~n_{c,0}n_{d,0}\,,
\end{align}
indicate the thermal average of annihilation cross-section. The total DM relic density is the sum of the individual 
relic densities, derived from the solution of the cBEQ in eq\,.~\ref{eq:cbeq_model}, to provide, 
\begin{align}
{\rm \Omega_{DM}h^2} = 2.744 \times 10^8 \left[m_{\phi}Y^{}_{\phi}+m_{X}Y_{X}^{}\right]_{x\to \infty}\,.
\end{align}
The free parameters in this model that crucially dictates DM phenomenology are,
\begin{align} 
\{m_{\phi}^{},~m_{X}^{},~m_{h_2}^{},~g_X^{},~\lambda_{\phi H}^{},~\lambda_{\phi S}^{},~\sin\vartheta\}\,.
\end{align}
The parameter space which is scanned here is
: $\{10^{-3}<\lambda_{\phi S}<1,~10^{-4}<\lambda_{\phi H}<10^{-1},~10^{-3}<g_{X}<10^{-1},~10^{-4}<|\sin\vartheta|<10^{-2},~1<m_{X}<550,~1<m_{h_2}<10^3,~1<m_{\phi}<10^3\}$ where masses are in GeV unit. We have calculated the DM relic densities and the spin-independent (SI) DM-nucleon inelastic scattering cross-section using MicrOMEGAs 6.0 \cite{Alguero:2023zol}, and the findings are presented and discussed below.
\begin{figure*}[htb!]
\centering
\includegraphics[width=0.475\linewidth]{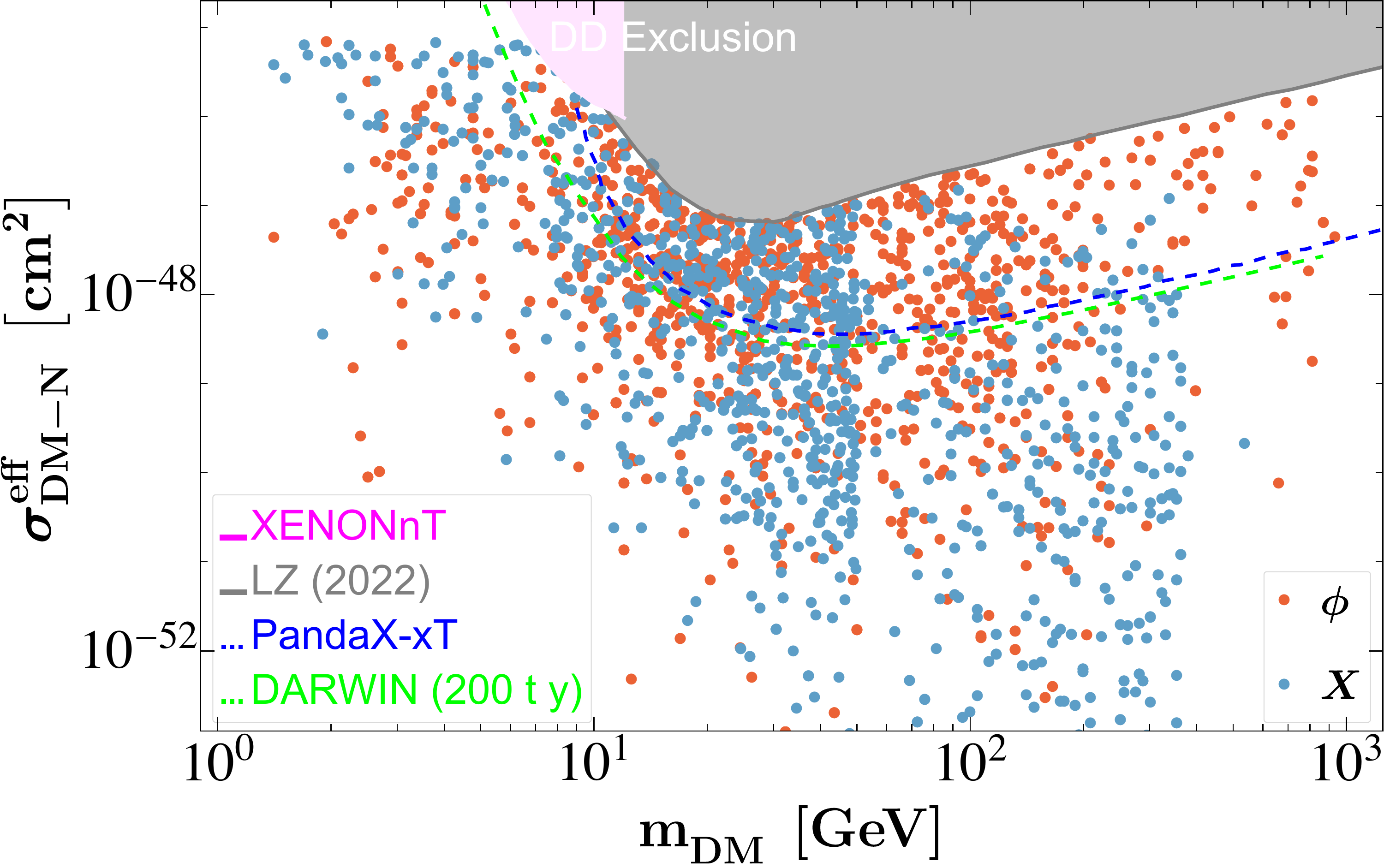}~~
\includegraphics[width=0.475\linewidth]{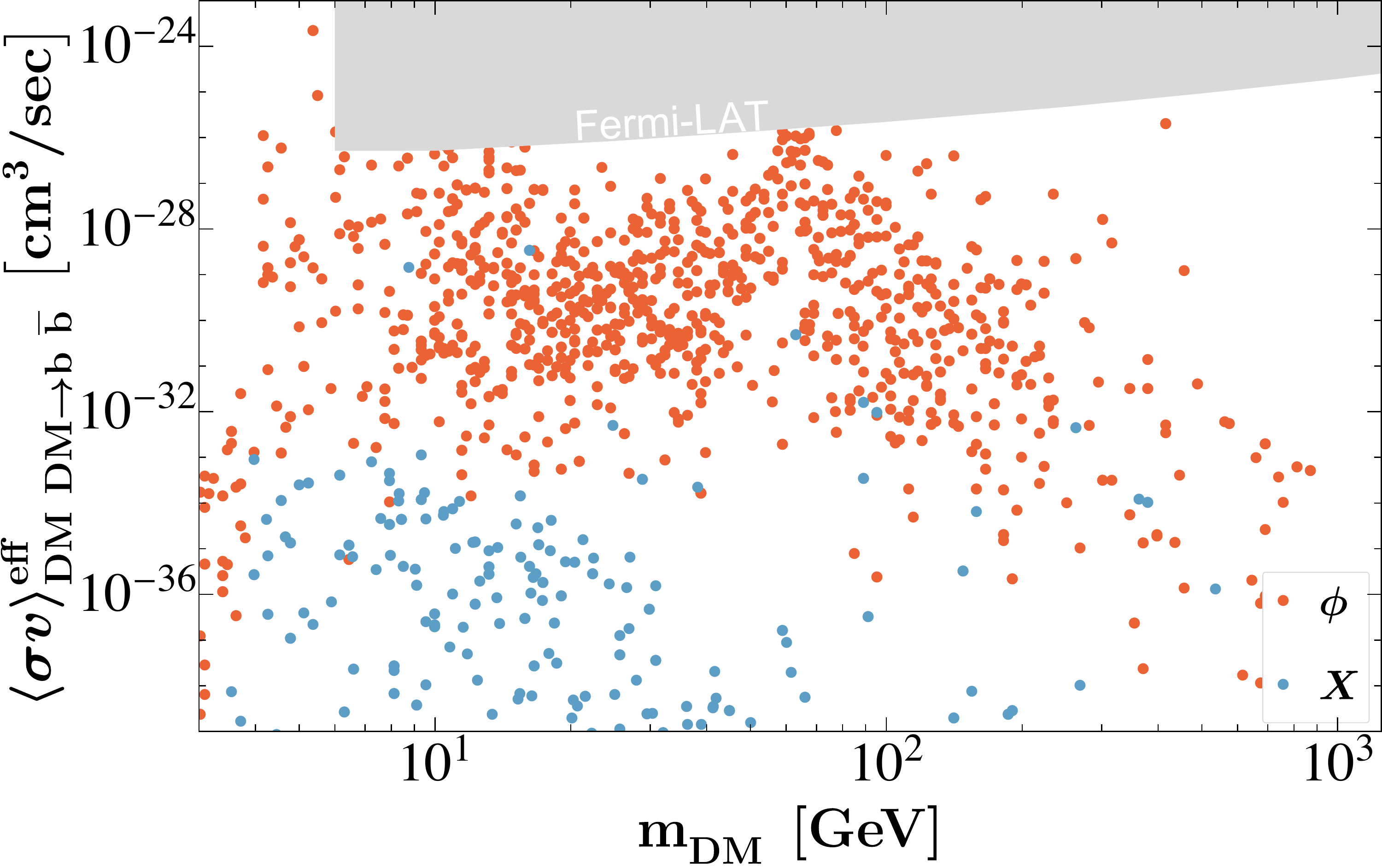}
\caption{{\it Left:} Allowed parameter space of the model in the \( \rm m_{\rm DM} - \sigma^{ eff}_{ DM-N} \) plane ($\phi$ in blue, $X$ in red points) 
that satisfies \( \mathcal{B}_{h_1 \to \rm inv} \leq 0.113 \) and \( \tau_{h_2} < 1~\text{sec} \), relic abundance and indirect  search ($\rm DM~DM\to b~\overline{b}$) constraints from Fermi-LAT and CMB. The bounds and future sensitivities from
experiments like XENONnT, LZ, PandaX-xT, DARWIN are shown. {\it Right:} points allowed by relic density, and recent direct detection constraint from LZ-2022 in 
${\rm m_{DM}}-\langle\sigma v\rangle^{\rm eff}_{\rm DM~DM\to b~\overline{b}}$ plane are shown with Fermi-LAT bound.}
\label{fig:relic-dd-id}
\end{figure*}

In the left plot of fig\,.~\ref{fig:relic-dd-id}, we illustrate the relic density allowed points in the $\rm m_{DM}-\sigma_{\rm DM-N}^{\rm eff}$ plane. The orange points represent real scalar DM ($\phi$), while the light blue points correspond to VBDM ($X$). The effective SI DM-nucleon scattering cross-section is defined as $\sigma_{i\rm N}^{\rm eff}=(\Omega_i/\Omega_{\rm tot})\sigma_{i\rm N}^{\rm SI}$, where the individual direct search cross-section is scaled by the ratio of its relative abundance with respect to the total one. 
The grey (LZ-2022 \cite{LZ:2022lsv}) and pink (XENONnT \cite{XENON:2024hup}) shaded regions indicate the DD exclusion limits, while the projected limits are shown by the dashed blue (PandaX-xT \cite{PandaX:2024oxq}) and green (DARWIN-200 t y \cite{Baudis:2023pzu, Baudis:2024jnk}) lines. The points scanned here
already obey indirect detection and collider search limits. Additionally, two peaks in the indirect detection signal $\langle \sigma v \rangle_{\rm DM~DM\to b~\overline{b}}$ occur near the $h_1$ and $h_2$ resonances, leading to the exclusion of some points in these regions. The major stumble block in coming up with a realistic model that produces a curvature 
in the direct search signal is that one of the DM components must have small mass, therefore susceptible to the invisible branching ratio of either the Higgs or $Z$ boson or 
whatever the DM-SM portal is. In our case, we need to adhere to the Higgs invisible decay bound, and we can address that by choosing a very small mixing angle 
$|\sin\vartheta| \lesssim 10^{-2}$. This was one of the reasons of having the second light scalar present in the scenario. From the figure, it is clear that there is plenty of 
breathing space for the model to survive the non-observation of direct search limit after addressing the observed relic density, with both the DM components providing detectable 
points in the vicinity of the exclusion limit and beyond.  Notably, we are considering spin independent direct search cross-section and the corresponding limits as obtained 
by the DM interactions in this model. 

In the right plot in fig\,.~\ref{fig:relic-dd-id}, we have shown the relic density allowed points in ${\rm m_{DM}}-\langle\sigma v\rangle^{\rm eff}_{\rm DM~DM\to b~\overline{b}}$ plane, where the colour coding remains the same as the left plot. All the points shown here are also allowed by direct and collider search constraints. 
This particular annihilation channel to the bottom quark pair is relevant as some points are excluded near the Higgs resonance region from the 
Fermi-LAT \cite{Fermi-LAT:2015att, Fermi-LAT:2016afa, Fermi-LAT:2015kyq} limit. 
There are points below the Higgs mass, which also get restricted because of the resonance enhancement due to the presence of the light mediator $h_2$.

\section{Two component dark matter signal in direct detection}\label{sec:2dm-dd}
After finding out the allowed parameter space of the model from the existing bounds, we now turn to the direct search observability of two DM components. 
Its worth repeating that we wish to focus on the direct search signal where a kink or a curvature in the recoil energy spectrum is observed 
\cite{Herrero-Garcia:2017vrl, Herrero-Garcia:2018qnz}. Absent any background, this indicates the presence of more than one DM components. In the context of the 
model presented above, it is therefore the exploration of the region of parameter space where the presence of two DM components can be realised.

Let us now turn to the formalism a little. The total observed differential event rate is the sum of the individual event rates from each DM component 
interacting with the target nuclei. The DM relic densities as obtained from the solution of cBEQ are given by,
\begin{align}
\Omega_X h^2=\dfrac{\rho_X}{\rho_{\rm crit}}\,, \quad \Omega_{\phi} h^2=\dfrac{\rho_{\phi}}{\rho_{\rm crit}}\,, \quad {\rm with} \quad \rho_{\rm loc}=\rho_X+\rho_{\phi}\,,
\end{align}
where the critical density is given by $\rho_{\rm crit}=\dfrac{3H_0^2}{8\pi G}\sim $, with $\rm H_0$ and $G$ representing the Hubble and gravitational constants respectively. 
Following the above relations, we can evaluate the DM densities in terms of relic densities,
\begin{align}
\rho_{\phi}=\rho_{\rm loc}\dfrac{\Omega_{\phi}h^2}{\Omega_{\phi}h^2+\Omega_Xh^2}\,,\quad{\rm and}\quad
\rho_{X}=\rho_{\rm loc}\dfrac{\Omega_{X}h^2}{\Omega_{\phi}h^2+\Omega_Xh^2}\,,
\end{align}
and
\begin{align}
\rho_{\phi}\sigma_{\phi}^N=\rho_{\rm loc}\dfrac{\Omega_{\phi}h^2}{\Omega_{\phi}h^2+\Omega_Xh^2}\sigma_{\phi}^N\equiv \rho_{\rm loc}\sigma_{\phi-N}^{\rm eff}\,,\quad{\rm and}\quad
\rho_{X}\sigma_{X}^N=\rho_{\rm loc}\sigma_{X-N}^{\rm eff}\,.
\end{align}
The time-averaged event rate for recoil, typically measured in events/(kg keV day), for a detector with a target nucleus\footnote{In this article, we have considered the $\rm {}^{132}Xe_{54}$ as our detector nuclei. However, we can do a similar analysis for other nuclei such as germanium (EDELWEISS \cite{EDELWEISS:2020fxc}) and sodium (COSINE-100 \cite{COSINE-100:2024wji}) for the spin independent case, while fluorine (PICO \cite{PICO:2019vsc}) for the spin-dependent case, etc.} of mass $m_A$ and characterized by 
the standard notation $(A, Z)$, is given by the sum of the individual contributions,\begin{align}
R_A(E_R)=R^{(1)}_A(E_R) +R^{(2)}_A(E_R)=F^2_A(E_R)\sum_{\alpha=X,\phi}(A_{\alpha }^{\rm eff})^2\frac{\rho_{\rm \alpha}\sigma^p_{\alpha}}{2m_{\alpha}\mu_{\alpha p}^2}\eta_{\alpha }(v_{m}^{\alpha},m_A,t)\,,
\label{eq:recoil}
\end{align}
where the nuclear form factor of the detector element with the mass number $A$ is denoted by $F_A(E_R)$, where $E_R$ denotes recoil energy. 
For the spin independent (SI) form factor, we use the Helm parametrization \cite{Lewin:1995rx, Helm:1956zz}.
The velocity integral ($\eta_{\alpha }(v_{m}^{\alpha},m_A,t)$) is defined as,
\begin{align} 
\eta_{\alpha }(v_{m}^{\alpha},m_A,t)= \int_{v_{m,A}^{(\alpha)}}d^3v \dfrac{f_{\rm det}^{(\alpha)}({\boldsymbol{ v}},t)}{v},
\end{align}
where $v_{m, A}^{(\alpha)} = \sqrt{\dfrac{m_A E_R}{2 \mu_{\alpha A}^2}}$ is the minimum velocity required for a DM particle $\alpha$ to produce a nuclear recoil with energy $E_R$ (the detector's threshold) in a nucleus of mass $m_A$.
The velocity distribution of DM particles in the detector frame, $ f_{\rm det}^{(\alpha)}(\boldsymbol{v}, t) $, satisfies $ f_{\rm det}^{(\alpha)}(\boldsymbol{v}, t) \geq 0 $ and $ \int f_{\rm det}^{(\alpha)}(\boldsymbol{v}, t) d^3v = 1 $. The velocity distributions in the detector and galaxy rest frames are related further by a Galilean transformation: $ f_{\rm det}(\boldsymbol{v}, t) = f_{\rm gal}(\boldsymbol{v} + \boldsymbol{v}_e(t)) $, where $ \boldsymbol{v}_e(t) $ is the Earth's velocity in the galaxy's rest frame.
Here, we have used the Standard Halo Model, assuming a local DM density of $\rho_{\rm loc} \simeq 0.4~\rm GeV/cm^3$ and a Maxwelian velocity distribution \cite{Kavanagh:2020cvn, deSalas:2020hbh, Sofue:2020rnl},
\begin{align} 
f_{\rm gal}(v)=\dfrac{1}{(2\pi v^2_{\alpha})^{3/2}}{\rm exp}\left(-\dfrac{3}{2}\dfrac{v^2}{v^2_{\alpha}}\right)\,,
\end{align} 
with a cutoff at the escape velocity, \(v_{\rm esc} = 550~\rm km~s^{-1}\) \cite{Smith:2006ym, Kafle:2014xfa}. The velocity dispersion $v_{\alpha}$ 
depend on the DM masses within the model following,
\begin{align} 
v_{\alpha}=v_0\left(\overline{m}/m_{\alpha}\right)^{1/2}\,,
\end{align} 
where  the canonical value for the velocity dispersion is $v_0\sim 270~\rm km~s^{-1}$ \cite{Foot:2012cs, Staudt:2024tdq}, and
\begin{align} 
\overline{m}=\sum_{k=1,2}n_km_k/\sum_{l=1,2}n_l=\rho_{\rm loc}/(n_1+n_2)=\left(\sum\limits_{k=1,2}\Omega_kh^2\right)\left(\sum\limits_{l=1,2}\dfrac{\Omega_lh^2}{m_l}\right)^{-1}\,.
\end{align} 
The scattering cross-section of DM particle $\alpha$ with the protons at zero momentum transfer is denoted by $\sigma_{\alpha}^p$, while $\mu_{\alpha p}$ represents the reduced mass of the DM particle $\alpha$ and the proton. 
The nuclear form factor of the nucleus $j$ is denoted by $ F_j(E_R)$. The effective mass number of nucleus $j$, interacting with the DM particle $\alpha$, is $A^{\rm eff}_{\alpha}=Z+(A-Z)f^n_{\alpha}/f^p_{\alpha}$ with $f^{n,p}_{\alpha}$ representing the SI coupling strengths of the DM particle $\alpha$ to neutrons and protons respectively. For the SI form factors, we use the Helm parametrization in our numerical analysis \cite{Helm:1956zz, Lewin:1995rx}. 
\begin{table}[htb!]
\centering
\setlength{\tabcolsep}{4.5pt} 
\renewcommand{\arraystretch}{1.7} 
\begin{tabular}{|c|c|c|c|c|c|c|c|c|c|c|c|c|c|c|c|}
\hline
\rowcolor{black!15}Benchmarks & $m_{\phi}~ \left[\rm GeV\right]$ & $m_X~ \left[\rm GeV\right]$ & $\Omega_{\phi}h^2$ & $\Omega_{X}h^2$ & $\sigma_{\phi-N}^{\rm eff}~\rm \left[cm^2\right]$ & $\sigma_{X-N}^{\rm eff}~\rm \left[cm^2\right]$ \\
\rowcolor{red!10}A&109.65&8.77&0.0352&0.0851 & $2.00\times 10^{-47}$ & $1.98\times 10^{-47}$\\
\rowcolor{cyan!10}B&77.17&12.16&0.0264&0.0934 & $5.38\times 10^{-49}$ & $2.06\times 10^{-48}$\\
\rowcolor{violet!10}C&66.07&12.42&0.0005&0.11902 & $4.99\times 10^{-48}$ & $1.91\times 10^{-47}$\\
\rowcolor{blue!5}D&51.29&8.42& $2.8\times 10^{-5}$ & 0.1202&$3.95\times 10^{-49}$ & $4.55\times 10^{-48}$\\
\rowcolor{magenta!10}E&40.74&8.46& $2.4\times 10^{-5}$ & 0.1208&$1.18\times 10^{-48}$ & $2.09\times 10^{-47}$\\
\rowcolor{lime!10}F&199.53&13.18&0.0379&0.0822 & $3.23\times 10^{-48}$ & $4.41\times 10^{-48}$\\
\rowcolor{green!13}G&83.18&4.27&0.0294&0.0915 & $1.10\times 10^{-47}$ & $1.10\times 10^{-47}$\\\hline
\end{tabular}
\caption{The benchmark points noted above are capable of displaying a noticeable kink in the recoil rate spectrum while also meeting the constraints from relic density, DD, ID, and collider searches of DM.}
\label{tab:kink}
\end{table}

In tab\,.~\ref{tab:kink}, we present several benchmark points that account for the current DM relic density, stringent direct detection limits from LZ (2022), indirect detection constraints on DM annihilation into bottom quarks, and collider limits on the Higgs invisible decay. We also have accounted for the limits on the mixing angle $\sin\vartheta$ and the BBN limit on the $h_2$ lifetime. Each benchmark, labeled A to G, shows a noticeable kink in the recoil energy spectrum following the methodology described above.
\begin{figure}[htb!]
\centering
\includegraphics[width=0.475\linewidth]{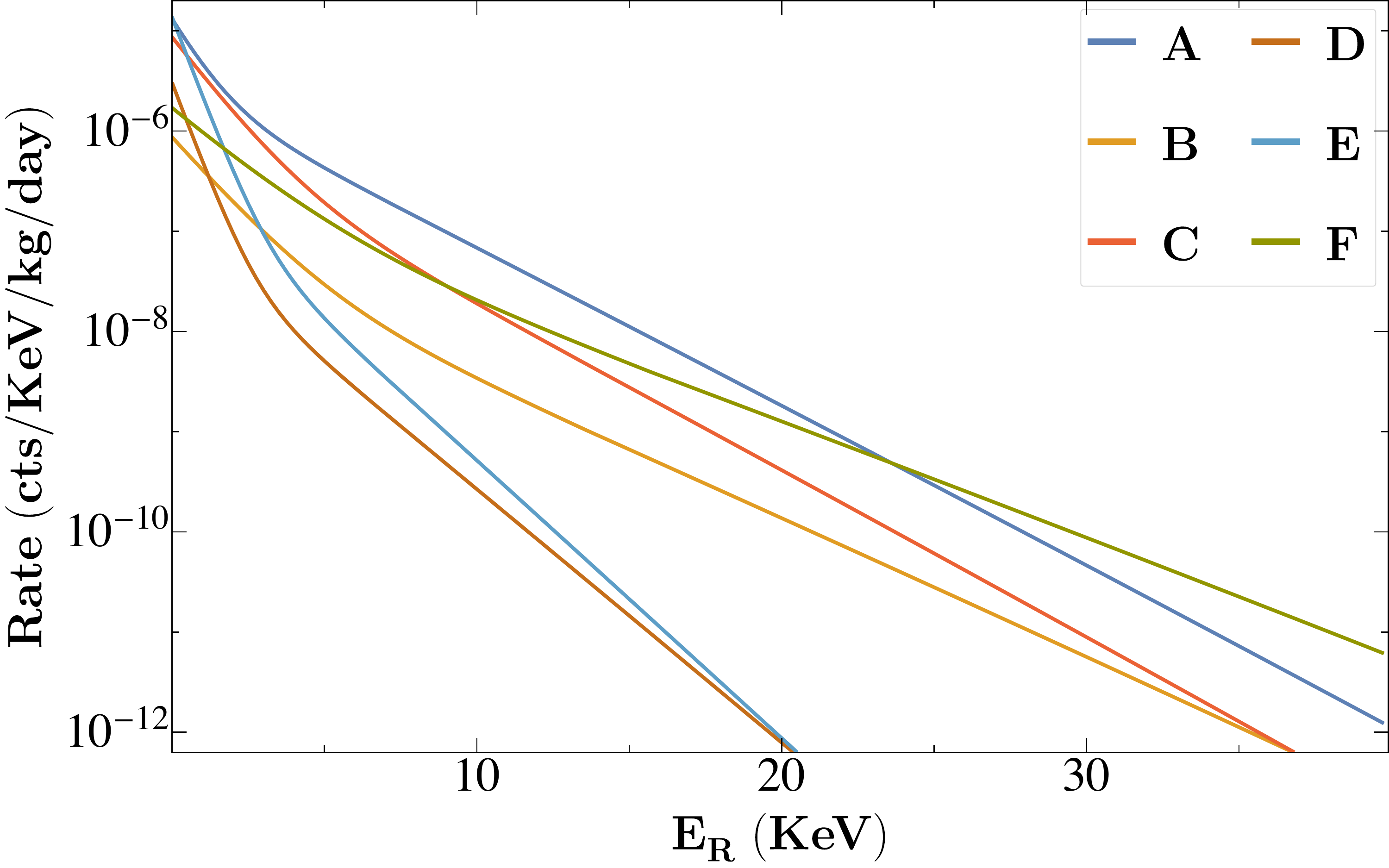}~~
\includegraphics[width=0.475\linewidth]{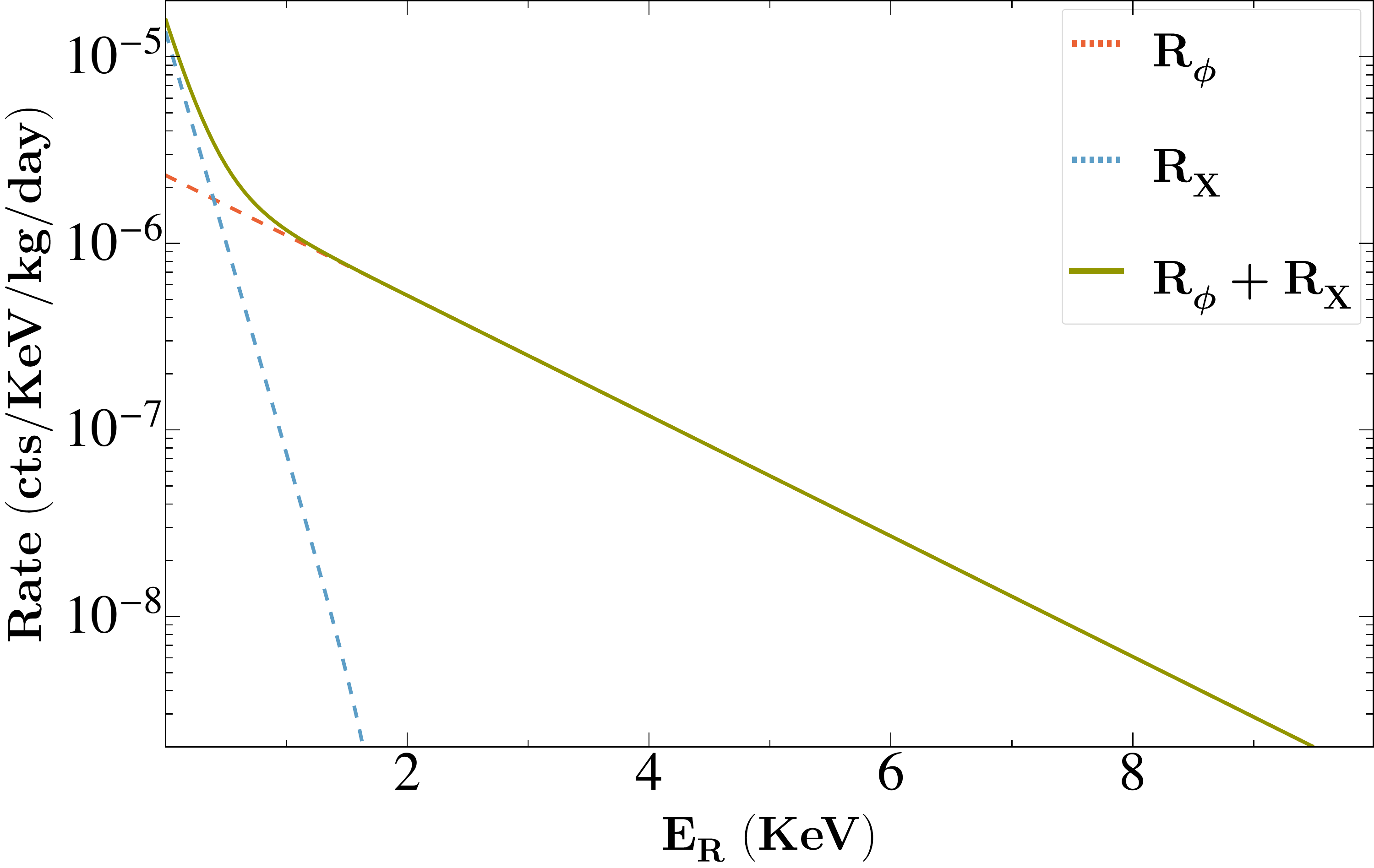}
\caption{Recoil rate spectrum for the benchmark points as in table \ref{tab:kink}. We show the variation of total event rate ($R_{\phi}+R_{X}$) with the recoil energy ($E_R$). 
The right plot shows benchmark G, while the left plot demonstrates benchmark points  A to F from table \ref{tab:kink}. The dashed blue and orange lines on the right panel correspond to the individual rates from vector boson and real scalar DM respectively.}
\label{fig:kink}
\end{figure}
The variation of the total time-averaged event rate as a function of recoil energy ($E_R$) is shown in fig\,.~\ref{fig:kink}. 
All the benchmark points described in table ~\ref{tab:kink} are shown here by different coloured lines. The kink typically appears around $E_R \lesssim 10$ keV. 
Its position depends not only on the DM mass but also on the relic density and the SI DM-nucleon scattering cross-section. In the left panel, $E_R$ ranges up to 40 keV, while in the right panel, it is limited to 10 keV for a closer inspection of the kink. We also show on the right panel, how the different slopes in $R_{\phi}$  and $R_{X}$ adds to the kink. It is 
obvious that there are regions of parameter space, where the slope for both DM components are same in the recoil spectrum, where the distinguishability is submerged. So it is 
legitimate to ask, what is the parameter that crucially governs the kink. The answer to that is DM mass. So, the larger the mass is, the smaller the slope is. Whenever, there is a 
substantial gap in the slope between the DM components, having similar direct search cross-section, such distinctive curvature will appear to indicate the presence of two 
DM components.   

The angles between the linear fit of the vector and scalar DM event rate curves, measured from the positive $E_R$ axis, are represented as $ \theta_{X} $ and $ \theta_{\phi}$, respectively. The resultant angle between these two curves is defined as $\theta = \theta_{X} - \theta_{\phi}$. If $m_X > m_{\phi}$, then $\theta_{\phi} < \theta_X$, resulting in a positive angle. Conversely, if $m_X < m_{\phi}$, then $\theta_{\phi} > \theta_X$, yielding a negative angle. However, other parameters, such as DM relic density and DD cross-section, can influence the slope, though their effects are mild and flexible, with a considerable impact on event rates.

\begin{figure}[htb!]
\centering
\includegraphics[width=0.65\linewidth]{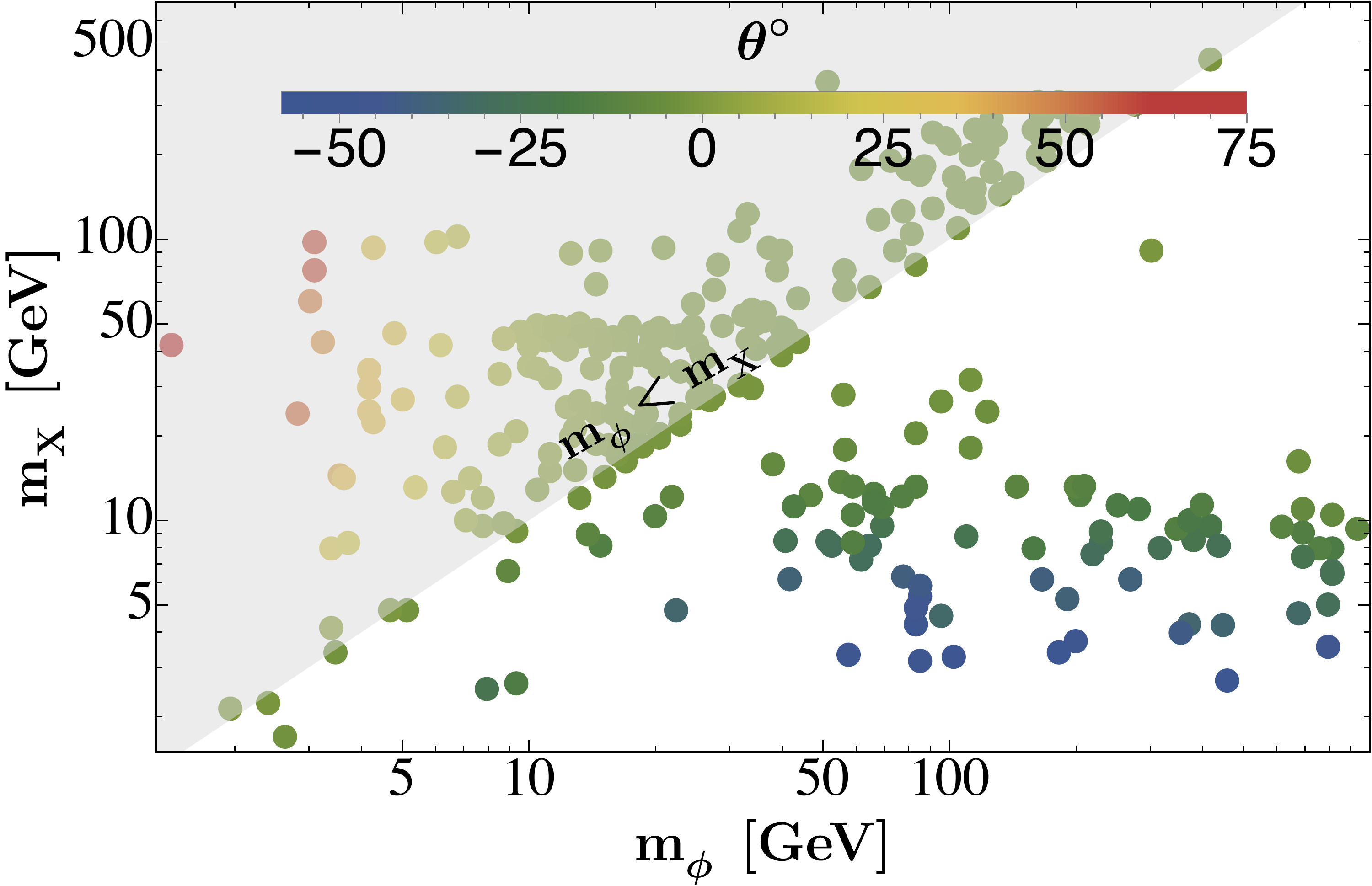}
\caption{Rainbow color bar shows the variation of angle between the two event rate lines w.r.t $E_R$, while the $\theta$ is in degree. The grey-shaded region is insensitive for experiments to discriminate the slope due to the difference in even rates, although accommodates large $\theta$; for details, see the main text.}
\label{fig:theta}
\end{figure}

Finally, in fig\,.~\ref{fig:theta}, we show the points in the $m_{\phi}-m_X$ plane that satisfy the DM relic density and respect the DD, ID, and collider constraints. Importantly, 
the rainbow color bar represents the angle ($\theta$, in degrees) between the two linear fit lines for the event rates of the two DM particles, $X$ and $\phi$ in the recoil energy 
spectrum. We see the presence of bluish or yellow-reddish points, where the angle is large to make the kink visible, 
spanning at large $m_\phi$ with small $m_X$. This also validates the claim made in model independent way in the 
references \cite{Herrero-Garcia:2017vrl, Herrero-Garcia:2018qnz} that the kink appears only when one mass is around $10~\rm GeV$, and the other is $\gtrsim 40~\rm GeV$. 
On the contrary, for the opposite mass hierarchy, i.e. $m_X>m_\phi$, energy densities are widely different with $\rm \rho_{\phi} \ll \rho_X $, 
and the event rates for $\phi$ is way smaller as 
$\rm R_{\phi} \ll R_X $ across the entire range of $ \rm E_R $. Consequently, this hierarchy is less promising for discriminating between the two DM components, although a finite slope difference exists between the two linear-fitted event rate curves. However, the situation alters if we can have $\rho_{\phi}^{} > \rho_X^{}$ for $m_{\phi}^{} < m_X^{}$ by incorporating new degrees of freedom into our model.
Thus, our work provides an example of a UV-complete vector-scalar model, where the distinguishability in direct search could be demonstrated. 
This is possible due to the presence of light scalar ($h_2$), which helps keep the VBDM under relic, even below the Higgs resonance. We also note that as a result, the benchmark points where VBDM is light, provides the majority of the relic density contribution. However, the possibility 
of having equal share of relic density is difficult, at least in this model, to provide a distinctive direct search signal for the presence of both.  

A statistical method using frequentist statistics for assessing the sensitivity of future experiments to distinguish between a single- and two-component DM scenario was discussed in detail by \cite{Herrero-Garcia:2017vrl}. It involves hypothesis testing and parameter estimation, where DM masses, the DM-nucleon scattering cross-section, and the energy density of each component and the same velocity dispersion assuming SHM are assumed free. In contrast, these parameters excepting for the DM mass are evaluated in our model following their interactions, adhering to all observation constraints. 
\section{Summary and conclusions}\label{sec:summary}
There are many possibilities embedded in multi-component DM, both in construction and phenomenology. 
The limits on the self-interaction of DM from the galaxy cluster observations like Bullet, Abell 
clusters does not clarify the ambiguity of its component type, as the self-interaction could be among the same or different kinds of particles that constitute the $26.8\%$ dark 
sector of the observed universe. An extended dark sector provides different phenomenological advantages specifically due to the DM-DM conversion, however an observational 
signal can only testify such hypothesis. Therefore, observation of two component DM signals in direct detection (DD), indirect detection (ID), and collider experiments are important to study. In this article, we have focused on multi-component DM signals in direct detection experiments. 

In model independent analysis \cite{Herrero-Garcia:2017vrl, Herrero-Garcia:2018qnz}, it was argued that a kink in the nuclear recoil rate can provide such a useful hint of having 
two different DMs coexisting and producing direct search signal. However, we don't know of an analysis where a UV complete model was studied to discuss such 
possibilities. Now, this is important as several constraints both from theoretical consistency as well as from observations like collider searches, indirect searches 
limit such models heavily with a potential possibility that the region of the parameter space where such distinguishability is observed, is actually discarded. 
This was the main point of  concern and rationale behind our study. Many of the simple kind of two component DM models studied so far fails to provide such distinguishability, 
where the model is valid, like having two scalar/fermion singlets; one singlet, one doublet etc. 

In the proposed model, we extend the SM by introducing a SM gauge-singlet real scalar $(\phi)$ stable under $\mathcal{Z}_2^{\prime}$ and a complex scalar $S$ respecting $U(1)_X$ gauge symmetry. The associated gauge boson $X$, stable under $\mathcal{Z}_2$, serves as the second DM candidate. The singlet-doublet mixing generates a new scalar $(h_2)$ together with the SM Higgs $(h_1)$. The mixing angle $(\vartheta)$ serves as a crucial variable, that can make one DM light, after addressing Higgs 
invisible decay and (in)direct detection of DMs, by choosing 
$|\sin\vartheta|\lesssim10^{-2}$. The individual relic densities are calculated by solving the cBEQ. We assume the light scalar $h_2$ to be in thermal bath, the rationale of which is 
explained in Appendix \ref{sec:equib}. Another strong constraint comes from the BBN, as the presence of a light scalar can alter the BBN prediction of the $\Lambda$CDM model. Therefore, the lifetime of this light scalar should be smaller than $1~\rm sec$. Further, for direct and indirect detection, we calculate the `effective' DM-nucleon scattering and annihilation cross-sections, as they most often contribute to total relic density unequally.

Further, we calculate the time-averaged event rate, assuming that both DMs scatter off the detector nuclei and making it recoil. 
We plot the time-averaged event rate with the recoil energy for several benchmark points, respecting all relevant constraints to show that a kink in the recoil rate can be observed.
The event rate depends not only on DM masses, DM number densities, DM-nucleon scattering cross-section, but also on the detector materials, and 
their sensitivity decides the minimum recoil energy required to produce detectable signals.
Our analysis is based mainly on $\rm {}^{132}Xe_{54}$ material used in LZ, XENON, PandaX, and the threshold recoil energy for nuclear recoil events is 
$\rm1~ KeV$ as used for XENONnT experiments.
The curvature of the resultant event rate spectrum, which is the summation of event rates corresponding to the two DMs, provides the proof of the existence of more than one DM. Here, we measure the curvature by measuring the angle between two asymptotic lines corresponding to the event rates of two DMs. The position of the resultant kink depends upon 
several parameters, like DM mass, relative abundances, DM-nucleon scattering cross-section etc. 
The observability of the kink, or what is the minimum angle $(\theta_{\rm min})$ below which we are unable to differentiate between single-component 
and two-component DM frameworks is another question. This is discussed partly via statistical analysis in \cite{Herrero-Garcia:2017vrl}.
According to our model, the maximum achieved angle is $\sim 75^{\circ}$, where $m_{\phi} \lesssim 5\rm~GeV$ and $m_X \gtrsim \rm 40~GeV$, and $\sim -60^{\circ}$, where $m_X \lesssim 5\rm~GeV$ and $m_{\phi} \gtrsim \rm 40~GeV$. However, for $m_X>m_\phi$, the energy densities are highly different, $\rm \rho_{\phi} \ll \rho_X $, 
and the event rates for $\phi$ is way smaller than $X$, as $\rm R_{\phi} \ll R_X $ across the entire range of $ \rm E_R $, thus making the observability of the kink in the recoil rate nearly impossible. 

Therefore, the requirement that the recoil rates can't differ too much for the DM components by having DM-nucleon cross-section at the same ballpark, 
but the slope in recoil rate requires to be different by having different DM masses, favours a specific mass hierarchy $m_X<m_\phi$ for our model to show the kink.
The difference in DM mass results in different relic density contributions. Such a conclusion was not possible to draw from a 
model independent analysis.  Further, the possibility of having a kink in the direct search signal together with that of a double bump signal in the collider is an interesting 
question, or of their complementarity, if exists, which we plan to study later. \\

{\bf Acknowledgement:} SB would like to acknowledge Grant No. CRG/2019/004078 from SERB, Govt. of India. Although the tenure of the project has ended now, we started 
looking into such possibilities, as part of the project deliverables.  

\appendix
\section{Feynman diagrams for DM relic density}
\label{sec:feyn}
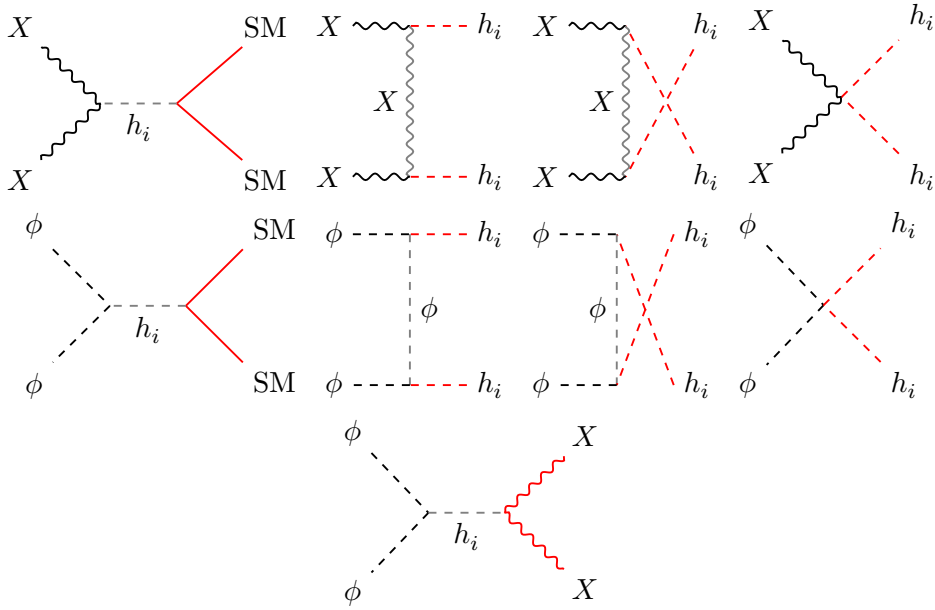
\begin{figure}[htb!]
\centering
\begin{tikzpicture}
\begin{feynman}
\vertex (a);
\vertex[right=1cm of a] (b);
\vertex[above left =0.75cm and 0.75cm of a] (a1){\(X\)};
\vertex[below left =0.75cm and 0.75cm of a] (a2){\(X\)};
\vertex[above right =0.75cm and 0.75cm of b] (b1){\(\rm SM\)};
\vertex[below right =0.75cm and 0.75cm of b] (b2){\(\rm SM\)};
\diagram*{
(a1) -- [line width=0.25mm,boson, arrow size=0.75pt,edge label={\(\rm \)},style=black] (a),
(a) -- [line width=0.25mm,boson, arrow size=0.75pt,edge label={\(\rm \)},style=black] (a2),
(b) -- [line width=0.25mm,plain, arrow size=0.75pt,style=red] (b1),
(b2) -- [line width=0.25mm,plain, arrow size=0.75pt,style=red] (b),
(b) -- [line width=0.25mm,scalar, arrow size=0.75pt,style=gray!50,style=black!50,edge label={\(\textcolor{black}{ h_{i}}\)}] (a)};
\end{feynman}
\end{tikzpicture}
\begin{tikzpicture}
\begin{feynman}
\vertex (a);
\vertex[below=2cm of a] (b);
\vertex[right =0.75cm and 0.75cm of a] (a1){\(h_{i}\)};
\vertex[left =0.75cm and 0.75cm of a] (a2){\(X\)};
\vertex[right =0.75cm and 0.75cm of b] (b1){\(h_{i}\)};
\vertex[left =0.75cm and 0.75cm of b] (b2){\(X\)};
\diagram*{
(a1) -- [line width=0.25mm,scalar, arrow size=0.75pt,edge label={\(\rm \)},style=red] (a),
(a) -- [line width=0.25mm,boson, arrow size=0.75pt,edge label={\(\rm \)},style=black] (a2),
(a) -- [line width=0.25mm,boson, arrow size=0.75pt,style=black!50,edge label'={\(\textcolor{black}{ X }\)}] (b),
(b1) -- [line width=0.25mm,scalar, arrow size=0.75pt,style=red] (b),
(b) -- [line width=0.25mm,boson, arrow size=0.75pt,style=black] (b2)};
\end{feynman}
\end{tikzpicture}
\begin{tikzpicture}
\begin{feynman}
\vertex (a);
\vertex[below=2cm of a] (b);
\vertex[right =0.75cm and 0.75cm of a] (a1){\(h_{i}\)};
\vertex[left =0.75cm and 0.75cm of a] (a2){\(X\)};
\vertex[right =0.75cm and 0.75cm of b] (b1){\(h_{i}\)};
\vertex[left =0.75cm and 0.75cm of b] (b2){\(X\)};
\diagram*{
(b1) -- [line width=0.25mm,scalar, arrow size=0.75pt,edge label={\(\rm \)},style=red] (a),
(a) -- [line width=0.25mm,boson, arrow size=0.75pt,edge label={\(\rm \)},style=black] (a2),
(a) -- [line width=0.25mm,boson, arrow size=0.75pt,style=black!50,edge label'={\(\textcolor{black}{ X }\)}] (b),
(a1) -- [line width=0.25mm,scalar, arrow size=0.75pt,style=red] (b),
(b) -- [line width=0.25mm,boson, arrow size=0.75pt,style=black] (b2)};
\end{feynman}
\end{tikzpicture}
\begin{tikzpicture}
\begin{feynman}
\vertex (a);
\vertex[above left =0.75cm and 0.75cm of a] (a1){\(X\)};
\vertex[below left =0.75cm and 0.75cm of a] (a2){\(X\)};
\vertex[above right =0.75cm and 0.75cm of a] (a3){\(h_{i}\)};
\vertex[below right =0.75cm and 0.75cm of a] (a4){\(h_{i}\)};
\diagram*{
(a1) -- [line width=0.25mm,boson, arrow size=0.75pt,edge label={\(\rm \)},style=black] (a),
(a) -- [line width=0.25mm,boson, arrow size=0.75pt,edge label={\(\rm \)},style=black] (a2),
(a) -- [line width=0.25mm,scalar, arrow size=0.75pt,style=red] (a3),
(a4)-- [line width=0.25mm,scalar, arrow size=0.75pt,style=red] (a)};
\end{feynman}
\end{tikzpicture}

\begin{tikzpicture}
\begin{feynman}
\vertex (a);
\vertex[right=1cm of a] (b);
\vertex[above left =0.75cm and 0.75cm of a] (a1);
\vertex[below left =0.75cm and 0.75cm of a] (a2);
\vertex[above right =0.75cm and 0.75cm of b] (b1);
\vertex[below right =0.75cm and 0.75cm of b] (b2);
\diagram*{
(a1) -- [line width=0.25mm,scalar, arrow size=0.75pt,edge label={\(\rm \)},style=black] (a),
(a) -- [line width=0.25mm,scalar, arrow size=0.75pt,edge label={\(\rm \)},style=black] (a2),
(b) -- [line width=0.25mm,plain, arrow size=0.75pt,style=red] (b1), 
(b2) -- [line width=0.25mm,plain, arrow size=0.75pt,style=red] (b),
(b) -- [line width=0.25mm,scalar, arrow size=0.75pt,style=gray!50,style=black!50,edge label={\(\textcolor{black}{ h_{i}}\)}] (a)};
\vertex[above left =0.75cm and 0.75cm of a] {\(\phi\)};
\vertex[below left =0.75cm and 0.75cm of a] {\(\phi\)};
\vertex[above right =0.75cm and 0.75cm of b] {\(\rm SM\)};
\vertex[below right =0.75cm and 0.75cm of b] {\(\rm SM\)};
\end{feynman}
\end{tikzpicture}
\begin{tikzpicture}
\begin{feynman}
\vertex (a);
\vertex[below=2cm of a] (b);
\vertex[right =0.75cm and 0.75cm of a] (a1);
\vertex[left =0.75cm and 0.75cm of a] (a2);
\vertex[right =0.75cm and 0.75cm of b] (b1);
\vertex[left =0.75cm and 0.75cm of b] (b2);
\diagram*{
(a1) -- [line width=0.25mm,scalar, arrow size=0.75pt,edge label={\(\rm \)},style=red] (a),
(a2) -- [line width=0.25mm,scalar, arrow size=0.75pt,edge label={\(\rm \)},style=black] (a),
(a) -- [line width=0.25mm,scalar, arrow size=0.75pt,style=black!50,edge label={\(\rm\textcolor{black}{ \phi }\)}] (b),
(b1) -- [line width=0.25mm,scalar, arrow size=0.75pt,style=red] (b),
(b) -- [line width=0.25mm,scalar, arrow size=0.75pt,style=black] (b2)};
\vertex[right =0.75cm and 0.75cm of a] {\(h_{i}\)};
\vertex[left =0.75cm and 0.75cm of a] {\(\phi\)};
\vertex[right =0.75cm and 0.75cm of b] {\(h_{i}\)};
\vertex[left =0.75cm and 0.75cm of b] {\(\phi\)};
\end{feynman}
\end{tikzpicture}
\begin{tikzpicture}
\begin{feynman}
\vertex (a);
\vertex[below=2cm of a] (b);
\vertex[right =0.75cm and 0.75cm of a] (a1);
\vertex[left =0.75cm and 0.75cm of a] (a2);
\vertex[right =0.75cm and 0.75cm of b] (b1);
\vertex[left =0.75cm and 0.75cm of b] (b2);
\diagram*{
(b1) -- [line width=0.25mm,scalar, arrow size=0.75pt,edge label={\(\rm \)},style=red] (a),
(a2) -- [line width=0.25mm,scalar, arrow size=0.75pt,edge label={\(\rm \)},style=black] (a),
(a) -- [line width=0.25mm,scalar, arrow size=0.75pt,style=black!50,edge label'={\(\rm\textcolor{black}{ \phi }\)}] (b),
(a1) -- [line width=0.25mm,scalar, arrow size=0.75pt,style=red] (b),
(b) -- [line width=0.25mm,scalar, arrow size=0.75pt,style=black] (b2)};
\vertex[right =0.75cm and 0.75cm of a] {\(h_{i}\)};
\vertex[left =0.75cm and 0.75cm of a] {\(\phi\)};
\vertex[right =0.75cm and 0.75cm of b] {\(h_{i}\)};
\vertex[left =0.75cm and 0.75cm of b] {\(\phi\)};
\end{feynman}
\end{tikzpicture}
\begin{tikzpicture}
\begin{feynman}
\vertex (a);
\vertex[above left =0.75cm and 0.75cm of a] (a1){\(\phi\)};
\vertex[below left =0.75cm and 0.75cm of a] (a2){\(\phi\)};
\vertex[above right =0.75cm and 0.75cm of a] (a3){\(h_{i}\)};
\vertex[below right =0.75cm and 0.75cm of a] (a4){\(h_{i}\)};
\diagram*{
(a1) -- [line width=0.25mm,scalar, arrow size=0.75pt,edge label={\(\rm \)},style=black] (a),
(a) -- [line width=0.25mm,scalar, arrow size=0.75pt,edge label={\(\rm \)},style=black] (a2),
(a) -- [line width=0.25mm,scalar, arrow size=0.75pt,style=red] (a3),
(a4)-- [line width=0.25mm,scalar, arrow size=0.75pt,style=red] (a)};
\end{feynman}
\end{tikzpicture}
\begin{tikzpicture}
\begin{feynman}
\vertex (a);
\vertex[right=1cm of a] (b);
\vertex[above left =0.75cm and 0.75cm of a] (a1){\(\phi\)};
\vertex[below left =0.75cm and 0.75cm of a] (a2){\(\phi\)};
\vertex[above right =0.75cm and 0.75cm of b] (b1){\(X\)};
\vertex[below right =0.75cm and 0.75cm of b] (b2){\(X\)};
\diagram*{
(a1) -- [line width=0.25mm,scalar, arrow size=0.75pt,edge label={\(\rm \)},style=black] (a),
(a) -- [line width=0.25mm,scalar, arrow size=0.75pt,edge label={\(\rm \)},style=black] (a2),
(a) -- [line width=0.25mm,scalar, arrow size=0.75pt,style=black!50,edge label'={\(\textcolor{black}{ h_{i}}\)}] (b),
(b) -- [line width=0.25mm,boson, arrow size=0.75pt,style=red] (b1), 
(b2) -- [line width=0.25mm,boson, arrow size=0.75pt,style=red] (b)};
\end{feynman}
\end{tikzpicture}
\caption{Feynman diagrams relevant for DM annihilation: $\rm \phi~\phi \rightarrow SM~SM$ and $\rm X~X \rightarrow SM~SM$, where ${\rm SM}\in \{h_{i},~\rm quarks,~leptons,~W^{\pm},~Z\}$, and $i=1,~2$. The bottom panel Feynman diagram shows the DM-DM conversion: $\phi~\phi\to\rm X~X$.}
\label{fig:feynman}
\end{figure}
In the two component DM model, constructed of a VBDM ($X$) and a scalar singlet ($\phi$) in thermal bath as described in Section\,.~\ref{mod:-vectro-scalar}, 
their annihilation cross-sections to the SM particles and conversion amongst themselves crucially decide the relic densities of the individual components. The Feynman graphs are 
shown in fig\,.~\ref{fig:feynman}. The individual relic densities also dictate the effective direct and indirect search cross-sections.  
\section{Thermal equilibrium for $h_2$}
\label{sec:equib}
\begin{figure}[htb!]
\centering
\includegraphics[width=0.475\linewidth]{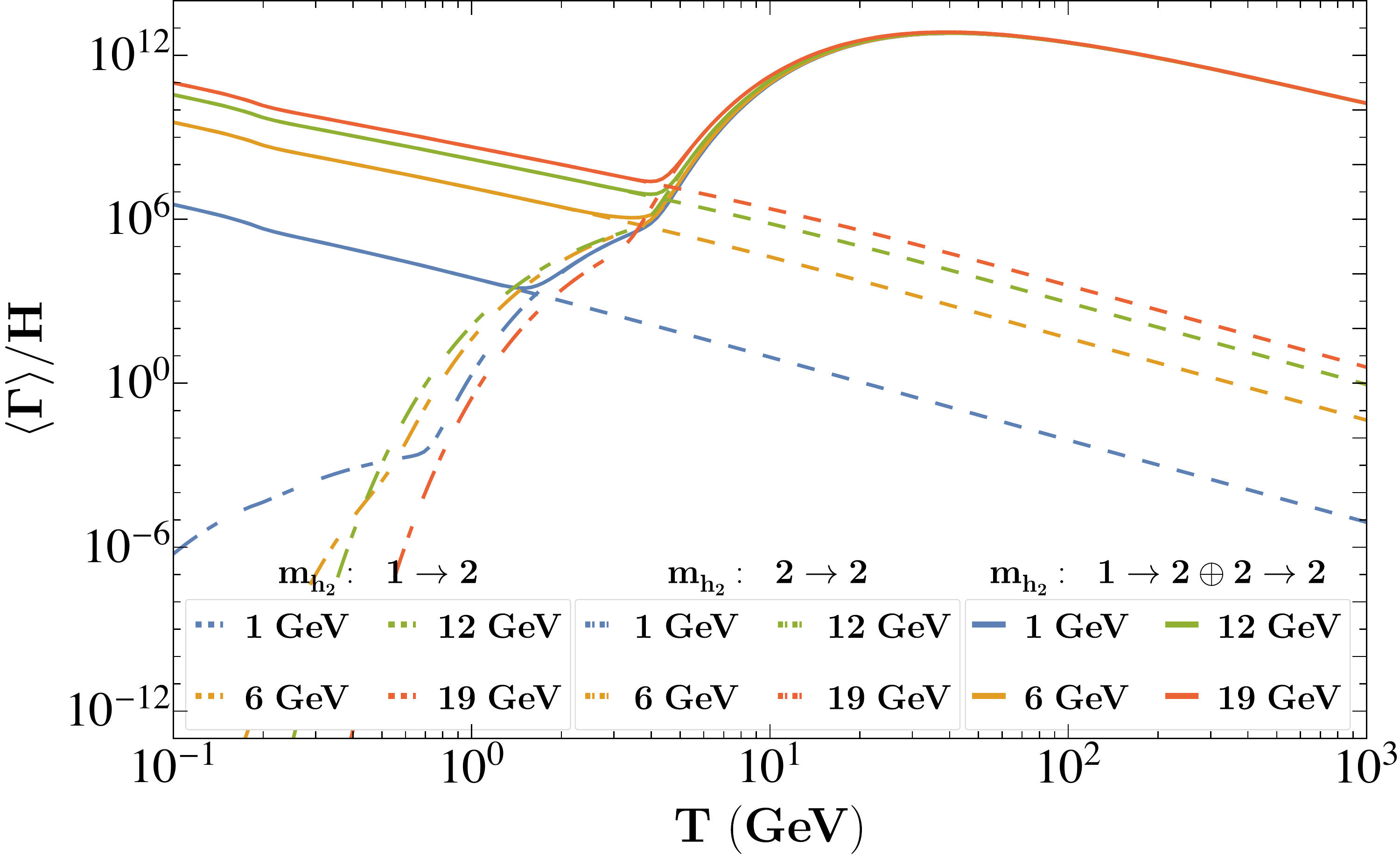}~~
\includegraphics[width=0.475\linewidth]{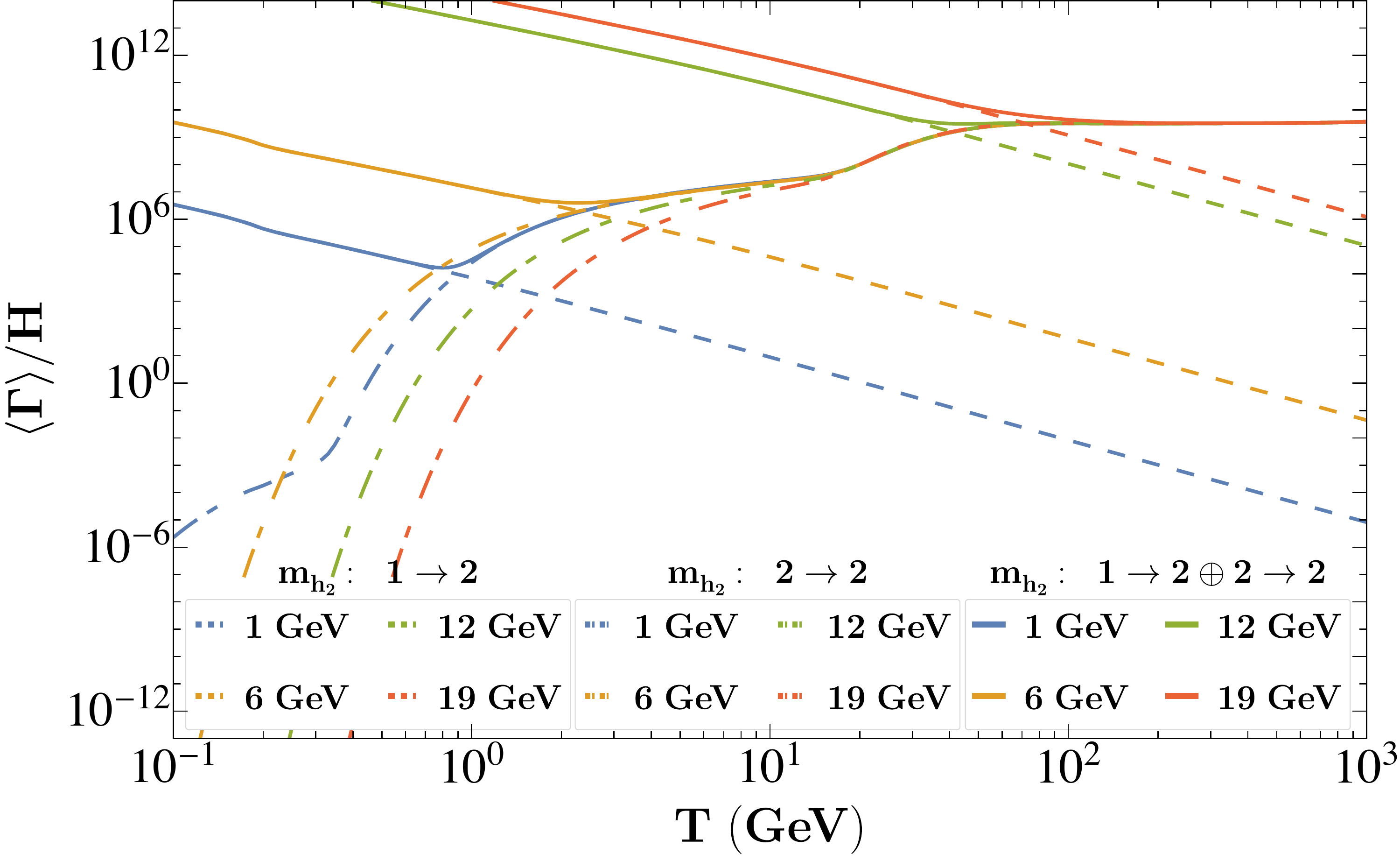}
\caption{$\langle \Gamma \rangle/H$ versus bath temperature $T$ plotted for $h_2$. The dashed, dot-dashed, and thick color lines show the interaction rate of $h_2$ for decay, scattering, and the combined decay plus scattering processes, respectively. The parameters kept constant are: $g_X=10^{-2},~\sin\vartheta=10^{-3},~\lambda_{\phi S}=10^{-1},~\lambda_{\phi H}=10^{-2}$, while for $\rm left: \{m_{\phi}=50~GeV,~{\rm and}~m_{X}=10~GeV\},~{\rm and~for~right:}~\{m_{\phi}=100~GeV,~{\rm and},~m_{X}=5~GeV\}$, with four distinct colors representing different values of $m_{h_2}^{}$.}
\label{fig:thermal-h2}
\end{figure}

The key assumption in solving the cBEQ for obtaining $\phi$ and $X$ relic densities, is that during DM freeze-out, $h_2$ remains in chemical and kinetic equilibrium, i.e., $n_{h_2}(T) = n_{h_2,0}^{}(T)$, and shares the same thermal bath temperature. The kinetic and chemical equilibria are maintained via scattering with relativistic SM fermions and annihilation and decay to the light SM fermions, respectively. The results are shown in fig\,.~\ref{fig:thermal-h2}, where $\langle \Gamma \rangle/H$ is plotted against bath temperature ($T$) for 
some allowed parameters of the model to show that in all cases including decay and scattering, $h_2$ remains in thermal bath. 
\newpage
\bibliographystyle{JHEP}
\bibliography{dd-2dm}
\end{document}